\documentclass[journal]{IEEEtran}

\ifCLASSINFOpdf
\else
\fi
\usepackage{array}
\usepackage{mdwmath}
\usepackage{multicol}
\usepackage[cmex10]{amsmath}
\usepackage{amsfonts}
\usepackage{array,color}
\usepackage{booktabs}
\usepackage{graphicx}
\usepackage{subfigure}
\usepackage{epstopdf}
\usepackage{cite}
\usepackage{algorithm}
\usepackage{algorithmic}
\usepackage{bm}
\usepackage{csquotes}
\usepackage{amssymb}

\usepackage{marvosym}

\usepackage{extarrows}

\begin{document}

\title{Distributed Cache Enabled V2X Network, Proposals, Research Trends and Challenging Issues}

\author{Di~Zhang,~\IEEEmembership{Member,~IEEE}
        \thanks{Di Zhang is with the School of Information Engineering, Zhengzhou University, Zhengzhou, 450001, China.}
}


\maketitle

\begin{abstract}
Nowadays, vehicle to everything (V2X) has been proposed to connect everything on the road. However, the current V2X network is motor-vehicle oriented. In contrast, the internet of sharing bicycle (IoSB) network is vastly and rapidly deployed as a feasible solution for the last mile problem (e.g., from station to home/office). To make full use of the devices for traffic safety and wireless communications, a comprehensive network model is needed. In this work, while investigating the existing studies, we propose a versatile V2X network with a distributed framework and heterogeneous caching method. In this V2X network, all the devices on the road (motor-vehicle, non-motor-vehicle, pedestrian, etc.) are connected. We further introduce a heterogeneous cache method for effective wireless transmission while utilizing the massive connected devices. The potential research trends on achieving high speed transmission, deep programming dedicated network slicing and big data/ machine learning (ML)/ edge computing based image recognition and reconstruction are highlighted to provide some insights for future studies. Finally, the challenging issues of path loss model,  channel model and ultra reliable and low latency communications (URLLC) of this V2X network are discussed.
\end{abstract}

%
\IEEEpeerreviewmaketitle

\section{Introduction}

In literature, the primary motivation of connecting vehicle is to prevent the collision\cite{Wu}. With this purpose, IEEE 802.11 p standard was released in 2010, and the dedicated short range communications (DSRC) with carrier frequency 5.9 GHz was allocated. In 2016, 3GPP release 14 has initialed the cellular-V2X (C-V2X), the 5G automotive association (5GAA) is established in the same year with the target on C-V2X solutions (i.e., vehicle to vehicle (V2V), vehicle to pedestrian (V2P), vehicle to infrastructure (V2I), vehicle to network (V2N)) for future mobility and transportation services). Based on the C-V2X network, opening issues of automatic piloting, traffic signal control, collision detection, emergency warning and optimal route planning have been raised\cite{Lu}. However, up to now, it is found that in V2X studies (DSRC and C-V2X), existing work mostly is motor-vehicle oriented, the other elements are playing the secondary roles for the vehicle's obstacle detection, collision prevention, etc. At the same time, the sharing bicycle system, with its flexible deployment and environment friendly characters, is becoming a convenient solution for the last mile problem (e.g., from station to office, home, etc.). In China, it is believed that the narrow band IoT (NB-IoT) based sharing bicycle system will become the first and largest IoT case in the coming years\cite{Huawei}. In order to accomplish the V2X's ambitious with connecting everything, fast speed transmission and intelligent traffic service, combination of current V2X, sharing bicycle and other devices is thus inevitable.

On the other hand, in wireless communications, content caching and sharing mechanism (CCSM) has been intensively studied. By invoking CCSM, content request can be satisfied from the temporary caches located in the core router, base station (BS) and neighboring user sides. The engaged transceiver equipments, links and resources (e.g., carrier frequency, consumed energy) can be greatly reduced via this method, which yields an improved system energy efficiency (EE) performance. The previous studies of content caching in vehicle network, however, were mostly restricted to the motor-vehicle V2V communications, for instance, \cite{Ahmed}. Due to a much smaller number of connected motor-vehicles, it is not an effective way to obtain requesting contents from the vehicle caches. The merits of content caching technology are reduced in this case.

In order to solve the aforementioned problems, here in this article, we propose the versatile distributed cache enabled V2X networks. All the devices (i.e., motor-vehicle, non-motor-vehicle, wearable smart device and cell-phone of the pedestrian) on-the-road are comprehensively connected for intelligent traffic system (ITS) control with regard to public and traffic safety. Besides, caches are distributed to all connected vehicles and devices (motor-vehicle, non-motor-vehicle, pedestrian, etc.), which are not limited to the motor-vehicles. We further exploit the massive connected vehicles and devices in the distributed cache enabled V2X networks for wireless transmission with the help of content caching technology. To effectively use the carrier frequency resources, we propose an initial co-existing solution for LTE, DSRC and millimeter wave (mmWave) frequencies. With the purpose of reducing carbon emission, the solar energy is invoked as well by introducing the solar energy converters to the vehicle bodies.

The future research trends and challenging issues on achieving the distributed cache enabled V2X networks are focused afterwards. For the implementation, a global perspective is needed to re-design the network architecture and infrastructure. We discuss the majority future research trends on achieving higher transmission rate, the effective and dedicate intelligent traffic control network architecture with networking slicing. The big data, machine learning (ML) and fog computing based image recognition and reconstruction is another topic for future research. The urban wireless path loss and channel models bring in great challenge for the quality of service (QoS) of wireless transmission and intelligence traffic control. To this end, we sketch the path loss model while incorporating prior studies. The ultra reliable communication and low latency requirement are the other challenging issues for its commercial use with regard to the public and traffic safety.

\section{The CCSM, V2X and Sharing Bicycle}
In this section, we discuss the existing work on content caching technology, V2X systems and sharing bicycle systems. In wireless communication, content caching mostly restricts to system EE analysis, optimal content division and positioning. The V2X and sharing bicycle systems, on the other hand, are taken as the promising IoT application scenes to reshape the future driving experiences and vehicle industry.

\subsection{The Features and Benefits of CCSM}

\begin{figure}[ht]
\centering
\graphicspath{{figure/}}
\includegraphics[width=3.4in]{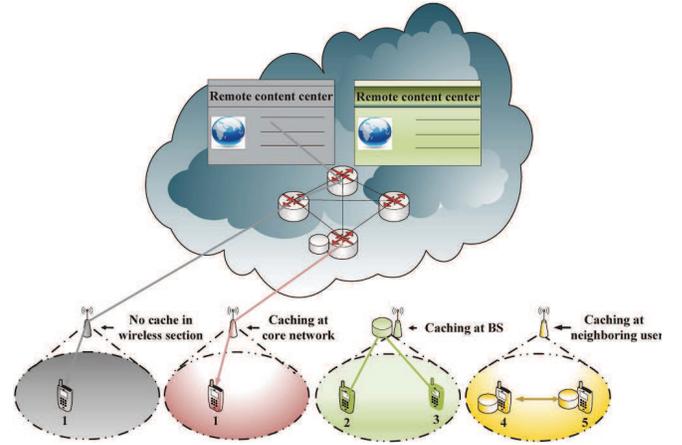}
\caption{The CCSM in wireless communications.}
\label{fig:CS}
\end{figure}

CCSM was initially originated from the information centric networking (ICN). In this paradigm, content is distributed in in-network storages with a name. Through multicast mechanisms, ICN can timely and efficiently deliver the contents to subsequent users. Nowadays, other than the upper layer studies, ICN is introduced to the wireless communications as well, which is known as the CCSM. As shown by Fig. \ref{fig:CS}, CCSM assumes that all the devices (core router, BS, mobile terminal, etc.) can cache and share the temporary contents for subsequent users. In this case, routing back and obtain the request contents from the remote center is not necessary. The benefits of CCSM can be summarized as follows:

\textbf{Alleviated network load}: The requesting data can be directly obtained from the distributed caches. Compared to routing back and retrieving from the remote content server, the core network load is alleviated. On condition that the subsequent user gets its requesting content from the neighboring user's caches, the BS network load is alleviated.

\textbf{Less engaged transceiver components}: CCSM needs less engaged transceiver components while providing the same amount of data to subsequent user. For instance, if the subsequent user is satisfied from the neighboring user's caches, the transceiver components of BS and core router are not needed; while from the BS caches, the core network transceiver components are not engaged; while from the core router caches, the transceiver components of subsequent routers are not necessary.

\textbf{Improved EE system performance}: The CCSM can reduce the transmission distance from transmitter to receiver. Because of the shorter distance, emission power is reduced, which brings less system energy consumption. Moreover, CCSM needs less engaged transceiver components. By keep those components into sleep mode, we may further reduce the energy consumption. In other words, CCSM consumes less energy while transmitting the same amount of contents. The system EE performance is therefore improved.

\subsection{The Current V2X and Sharing Bicycle Networks}

\begin{figure}
\centering
\graphicspath{{figure/}}
\includegraphics[width=3.5in]{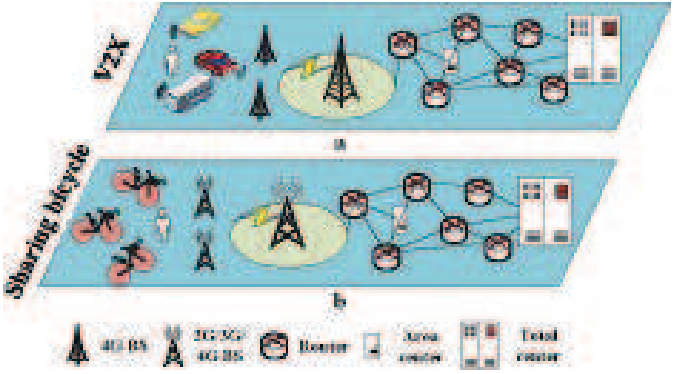}
\caption{The current V2X and sharing bicycle systems.}
\label{IoV}
\end{figure}

There are two types of V2X networks, i.e., DSRC and C-V2X. DSRC is to provide active safety communications contributes to safe driving while connecting the vehicles to RSU or other vehicles. In contrast, C-V2X aims to connect not only the vehicle, pedestrian, but also other elements (for instance, traffic lights, sensors) on the road. It can provide faster speed and lower latency transmission through the cellular network compared to DSRC. As shown by Fig. \ref{IoV}. a, vehicles can connect to the roadside units (RSU) via V2I, connect to the network via V2N, connect among each other via V2V and connect to the pedestrian via V2P. While communicating, DSRC relies on the wireless local area network (WLAN) physical transmission (PHY) and medium access control (MAC). Meanwhile, C-V2X works on two mode: 1) direct communication mode (DCM), for instance, V2V, V2P; and 2) network based communication mode (NBM), for instance, V2N, V2I. The benefits of V2X network can be summarized as follows.


\textbf{Safe and intelligent transport}: The centers can broadcast emergency warning message to vehicles via media    access  control (MAC) broadcasting to guide the vehicles for rescue and escape. With the shared real time information, the vehicle can adjust its travel speed to make the green light, take the optimal route to avoid traffic congestion, keep safe distance to avoid crash, etc.

\textbf{Improved in-car entertainment service}: The C-V2X networks enable the vehicle smart device communications by wireless connections working on the unlicensed spectrum (e.g., Bluetooth connection between vehicle and cellphone). For instance, the smart devices can lock or unlock the vehicle, play video or music through the vehicle's speakers. Besides, DCM and NBM working on the licensed spectrum makes the access of in-car high-definition media and virtual reality applications a reality.

Different from the C-V2X network, as shown by Fig. \ref{IoV}. b, the sharing bicycle network, mainly relies on the general packet radio service (GPRS)/Bluetooth/NB-IoT/global positioning system (GPS) technologies for nearest bicycle finding, lock/unlock, information sharing/uploading, location tracking, etc. That is, with the help of GPS and NB-IoT, potential user can find the nearest bicycle and unlock it via GPRS/Bluetooth/NB-IoT. While riding, the real time information (location, speed, acceleration, consumed energy, etc) will be recorded by the apps or transmitted to the remote center. The main benefits of the sharing bicycles network can be listed as follows.

\textbf{Last mile problem solutions and lower carbon emission}: Sharing bicycle network can tackle down the last mile problem (e.g., from bus/subway station to office/home). In this case, more people will prefer the public transport, especially when travelling within the urban area. Carbon emission thus is greatly reduced. Additionally, with less vehicles on the road, the sharing bicycle network can also alleviate the traffic congestion.

\textbf{Personal health diagnosis}: Sharing bicycle network can record the rider's personal data (e.g., calorie consumption, heart rate, blood pressure, ride distance) with apps or connected wearable sensors. Afterward, with some analytical tools, the rider can evaluate his (her) physical condition. The doctor may also access these data to provide professional health diagnosis under the rider's permission.

However, the current V2X networks and sharing bicycle networks are disconnected. We might cause a road accident when vehicles and bicycles cannot detect each other. On the other hand, we cannot provide the optimal routing without awareness of all types of devices on the road. It is also worth noting that in V2X networks, vehicle is playing a dominant role. The other elements are playing the secondary roles to assist its safe driving, automatic driving, in-car entertainment applications, etc.

\section{The Proposed Distributed Cache Enabled V2X Networks}

By incorporating the existing works on content caching, V2X and sharing bicycle, the proposed distributed cache enabled V2X networks are elaborated in this section.

\subsection{The Proposal}

\begin{figure}
\centering
\graphicspath{{figure/}}
\includegraphics[width=3.5in]{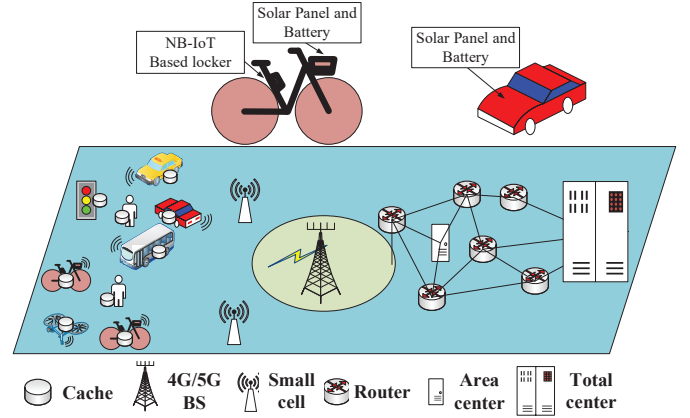}
\caption{The proposed distributed cache enabled V2X networks. All V2X devices on the road are connected for public and traffic safety with the CCSM. In addition, solar panel and battery are equipped in V2X devices as well.}
\label{fig:mix}
\end{figure}

The distributed cache enabled V2X networks can be shown by Fig. \ref{fig:mix}. Here in the proposed networks, CCSM is introduced to the core router, BS and all types of devices on the road. In reality, the cache can be integrated into the entertainment system of the car, and located in the intelligent locker system of the bicycle. In order to utilize the clean energy, we introduce solar energy converter to the V2X devices\footnote{While using the word devices, we mean all devices besides the vehicle, such as cell-phone, laptop, wearable smart devices.} (e.g., solar panel on vehicle and device body). Due to the bicycle's limited energy harvest ability, battery is invoked to equip within the basket. That is, the battery unit is equipped to the bottom of the basket, the solar energy converter is located over the battery unit. For the sharing bicycle within this system, we discard the GPRS/Bluetooth based locker while utilizing the NB-IoT based intelligent locker for fast lock/unlock. According to the report, lock/unlock can be completed within 1 s with NB-IoT, the payment process has been dropped from 25 seconds to less than 5 seconds, while battery life has been prolonged from 1 or 2 months to more than 2 years\cite{Huawei}.

In order to provide fast speed transmission for the connected devices, the distributed V2X networks can use 5G technologies in its NBM transmission. For implementation, we can put the massive multi-input-multi-output (MIMO) BS above the building, and vastly deploy the small cell BSs (femotocell, picocell) to roadside units. For the carrier frequency, NBM reuses the existing cellular frequency resources for cellular to V2X devices transmission, and the mmWave and DSRC frequencies are dedicated to the V2V, V2X, everything to vehicle (X2V), small cell to vehicle (SC2V) and small cell to pedestrian (SC2P) transmissions. Detailed comparison between the distributed cache enabled V2X networks and prior cache enabled vehicle networks, V2X, sharing bicycle are given by Table \ref{comparison}. As shown here, the proposed distributed cache enabled V2X networks comprehensively cover the existing technologies, which enables the ITS and wireless information delivery. The main features and benefits of the proposed networks, while comparing with prior work, can be listed as follows:

\begin{table*}[!t]
\centering
\caption{Comparison between prior work and the proposed distributed cache enabled V2X networks}
\label{comparison}
\begin{tabular}{|c|c|c|c|c|c|c|c|c|}
\hline
& Vehicle & Bicycle & Pedestrian & Content caching & Solar energy & DSRC & LTE & 5G \\ \hline
Prior cache enabled vehicle networks & ${\surd}$ &${\times}$ &${\times}$ & ${\surd}$ &${\times}$& ${\surd}$ & ${\surd}$ &${\times}$\\ \hline
Current V2X system & ${\surd}$ &${\times}$ &${\surd}$ &${\times}$ &${\times}$ & ${\surd}$ & ${\surd}$ & ${\surd}$ \\ \hline
Sharing bicycle system &${\times}$&${\surd}$ &${\times}$ &${\times}$& ${\surd}$ &${\times}$& ${\surd}$ &${\times}$\\ \hline
Our proposal & ${\surd}$ & ${\surd}$ & ${\surd}$ & ${\surd}$ & ${\surd}$ & ${\surd}$ & ${\surd}$ & ${\surd}$\\ \hline
\end{tabular}
\end{table*}

\begin{figure}[!t]
  \centering
  \graphicspath{{figure/}}
  \subfigure
  {
    \label{fig:pcr} 
    \includegraphics[width=1.65in]{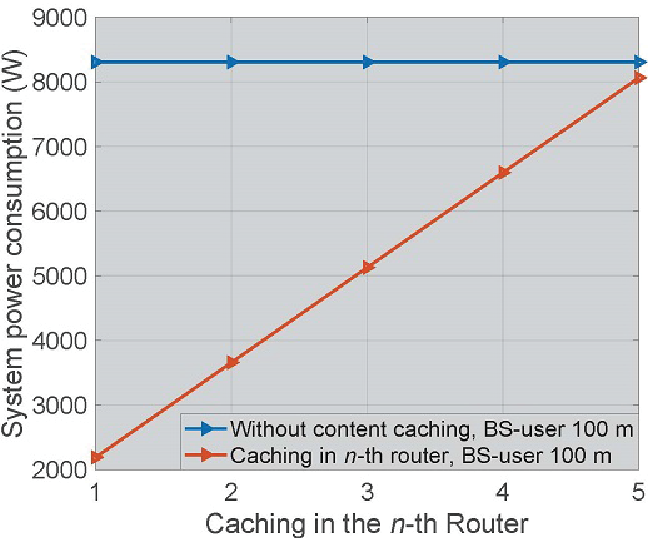}}
  \subfigure
  {
    \label{fig:epcr} 
    \includegraphics[width=1.65in]{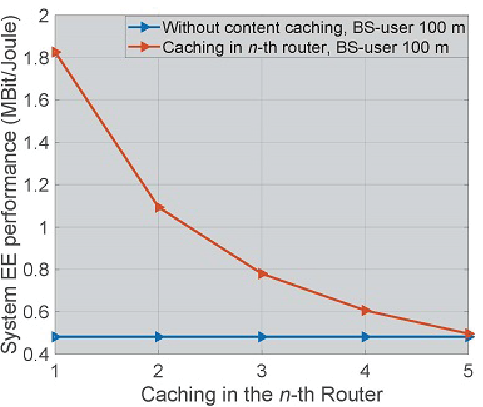}}
  \label{fig:pcrv} \\ 
   \subfigure
   {
    \label{fig:subfig:b} 
    \includegraphics[width=1.65in]{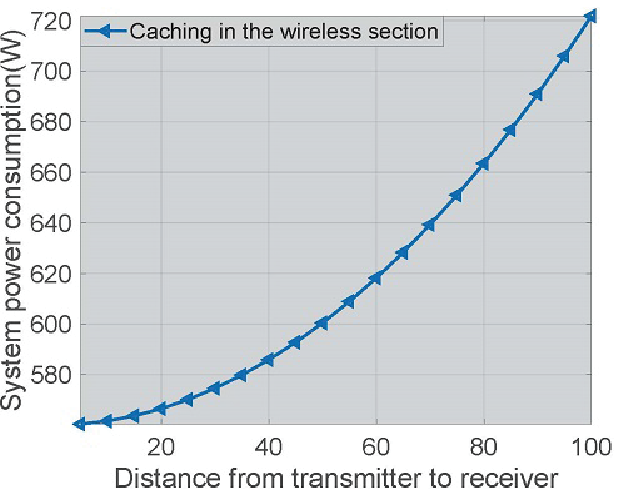}}
  \label{fig:subfig} 
   \subfigure
   {
    \label{fig:epcrv} 
    \includegraphics[width=1.65in]{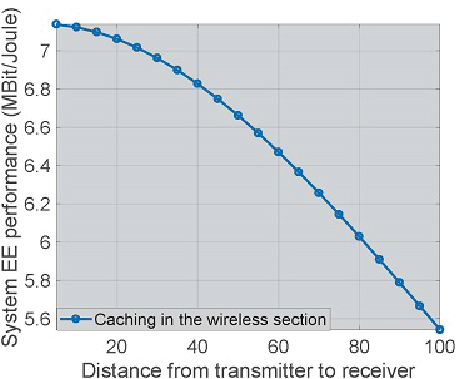}}
    \caption{The power consumption and system EE performance in different scenarios while transmitting 2G data within 1 second, carrier bandwidth 20MHz, by further considering the machine room (400 W), erbium doped fibre amplifier (EDFA, 243 W) and other core network elements. In upper two figures, caching in the core router and without caching is compared, where the bottom figures are system power consumption and EE performances of caching in the wireless section.}
      \label{fig:pe}
\end{figure}

\textbf{Improved intelligent traffic system and wireless transmission:} In this proposed distributed cache enabled V2X networks, all devices on the road are connected for public and traffic safety. The connected V2X devices are playing the equal roles for wireless communication. With the shared upload information from all vehicles and devices on the road, the control and service center can more effectively broadcast the emergency information and guide the vehicle (speed, accelerate, route, etc.) on the road. By sharing the information amongst cooperating vehicles, the automatics piloting system can be accomplished even without the help of center. Additionally, by adopting the 5G and CCSM, the proposed network can provide more efficient high speed wireless transmissions.

\textbf{Clean energy and more reasonable system model:} The V2X devices can use solar energy for communication, and store it in their batteries. With the clean solar energy, the distributed V2X networks can reduce the carbon emission and air pollution. Additionally, compared to prior cache enabled studies, e.g., \cite{D2DTWC2017}, solar energy's usage of the distributed V2X networks brings higher energy capacity to these V2X devices, which makes CCSM a more reasonable choice for V2X communications. For instance, with more energy, the V2X devices can increase its transmission power for faster speed transmission or prolong the transmission time.

\textbf{Reduced energy consumption and enhanced EE performance:} More energy can be saved while obtaining the request information from caches due to the shorter transmission distance and less engaged transceiver devices. The system EE performance is enhanced as well via this method. As shown by Fig. \ref{fig:pe}, content caching can greatly reduce the energy consumption of the distributed cache enabled V2X networks, especially by caching in the wireless section (BS, vehicle, device, etc.). This is mainly due to the reduced core network energy consumption. It is worth to note that giving the battery constraint, caching in the wireless section is only feasible when they have enough power.

\textbf{Usage of mixed frequency resources:} The DSRC 5.9 GHz frequency as well as mmWave frequency are dedicated for the wireless communications with V2V, V2P, V2I, V2N, SC2V, SC2P and P2P in the urban environment. This is mainly due to the high atmospheric attenuation and easily blocked features of higher frequency. By invoking DSRC and mmWave frequencies, the user can employ a wider carrier bandwidth for its large volume information transmission within a limited power under the assistance of content caching technology. In contract, to reuse the existing resources, LTE frequency is invoked for the BS to vehicle (BS2V) and BS to pedestrian (BS2P) transmissions.

\section{Future Research Trends and Challenging Issues}
In this section, we discuss the future research trends and challenging issues to accomplish the proposed cache enabled V2X networks. Methods on achieving the high speed transmission, flexible and dedicated network slicing service, big data and machine learning based image recognition and reconstruction, are main research topics for future research. Additionally, the accurate urban channel and propagation model, ultra reliable communication and low latency are big challenges of the distributed cache enabled V2X networks' implementation.
\subsection{Future Research Trends}

\subsubsection{Solutions for More Than 10 Folds Faster Transmission Rate}

In literature, the massive MIMO is raised as an essential element of 5G. It is proved that with antenna number growing, we can obtain more channel degree of freedom (DoF), which yields faster transmission rate and link reliability. On the other hand, mmWave recently is emerging as a vital element of 5G. According to Shannon theory, the achievable transmission rate can be boosted up while increasing the carrier frequency bandwidth value. Recently, the non-orthogonal multiple access (NOMA) is also intensively studied with regard to 5G's spectrum efficiency (SE). It allocates the same carrier frequency resource for multiple user's information transmission, whereas the encode and decode procedures are executed according to the allocated different power values (power domain NOMA) or codes (code domain NOMA)\cite{Zhang17}.

Meanwhile, the HetNets technology with co-existing macro cells and small cells are proposed. In HetNets, macro cell are used to provide wide coverage area, whereas in the cell edge areas, small cells are utilized to improve the connection and transmission quality. The carrier aggregation (CA) and coordinate multiple point transmission (CoMP) based cloud radio access network (C-RAN) can further leverage the transmission rate. Additionally, as talked before, content caching based network evolution is another interesting topic which attracts increasing attentions from both industry and academia. However, according to 5G Summit in Silicon Valley, scholars claimed that the 5G's higher transmission rate cannot be simply achieved with existing technologies up to now, redesigning the whole network architecture, combining the existing technologies, working on new air radio technologies are comprehensively needed with a joint force from both academia and industry.

\subsubsection{The Dedicated Network Slicing Service}
In prior wireless generations, for application scene with fast transmission rate, wide coverage area, ultra reliable communications, low latency communications, mostly a specialized network architecture should be established. Those specialized network architecture, once established, is hard to accommodate updates. Meanwhile, with large number of accessing vehicles and devices, the management of such a large scale network becomes troublesome\cite{HuangWC}. To cope with the diversity, dedicated network slicing technology\cite{Taleb} can be a feasible choice.

Network slicing technology is able to virtually divide the network into multiple co-existing sub-networks. It can adaptively assign the needed resources and establish the optimal router for each sub-network (network slicing service). Since the divisions and resource allocations are based on the software defined deeply programming, it can reuse the resources and update the existing sub-network(s) or establish a new dedicated sub-network once needed. With software defined network (SDN) controller and orchestration, we can control the connected vehicles and devices by creating a comprehensive network slice. In addition, it is also possible to set up different dedicated sub-networks for the high speed data transmission and intelligent traffic control scenarios, and adaptively adjust the allocated resources according to the update information. How to effectively tailor the dedicated network slice with given resources and flexibly adjust the sub-network to cater to the updates with network slicing technology will be an interesting topic for future study.

\subsubsection{Big Data, Machine Learning and Fog Computing Based Image Recognition and Reconstruction}

\begin{figure}[ht]
\centering
\graphicspath{{figure/}}
\includegraphics[width=3.4in]{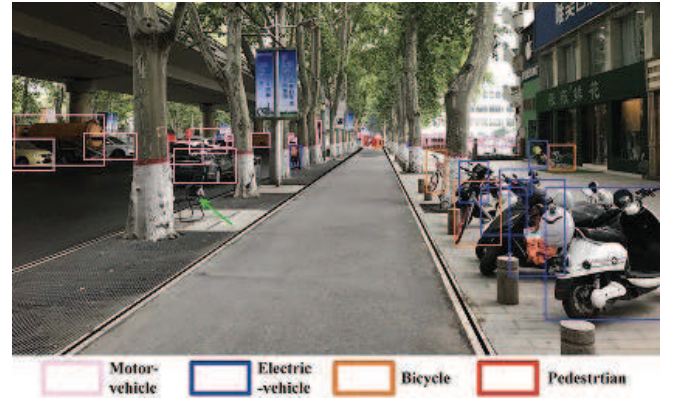}
\caption{An explanatory description of labeling data in image detection. The sharing bicycles and ordinary bicycles are all labeled as bicycle. Additionally, the chair is not labeled (marked with green arrow), which might bring risk to the vehicle while applying in automatic piloting for image detection in the following frames.}
\label{bdml}
\end{figure}

As known, automatics pilot is calling for real time ultra reliable image recognition and reconstruction. Due to the large scale and even fast growing data, previous iterative ML algorithm is not an effective way due to the space and time limitations. To accelerate the recognition speed, large scale deep learning (DL) and high-rank matrix factorization (MF) methods are proposed with massive parameters for improved filtering problems\cite{Xing15}. Yet the processing time and required resources are still huge. This is mainly due to the fact that while detecting the objectives, we need a set of labeled training data from a large scale data volume. This is the most time consuming process that called as the ground truth labeling\cite{nvidia}. On the contrary, if less parameters are considered, some elements on the road might be omitted, which will bring in risk for the traffic and public safety. For instance, while labeling the data in Fig. \ref{bdml}, the chair is not labeled (marked with green arrow, next to the first tree counting from the bottom left). For the image detection in the following frames based on the labeled data with Fig. \ref{bdml}, once the objective (unlabeled chair) was located on the motor-way, an accident might be caused. In addition, the sharing bicycle and ordinary bicycle are all recognized as the bicycle with same label to accelerate the recognition speed in the following steps.

Fortunately, Matlab has released the automated driving system toolbox to accelerate the labeling process, in addition, the Kanade Lucas Tomasi algorithm (KLT) can be invoked to label the objects in the first frame, and track them in the following frames\cite{nvidia}. The specialized GPU (for instance, NVIDIA a Tesla K40c) can be used to speed up the training process as well. However, with data scale growing, it is still a challenging and time-consuming task. To this end, a trade-off strategy between the engaged parameters (or labels) and real-time recognition and reconstruction requirement is needed. Additionally, a joint force of the off-the-shelf in-car cooperative edge computing chips and information sharing strategies amongst vehicles is needed. The developments on effective big data analysis and DL based recognition algorithms, cooperative edge and cloud computing strategies
, specialized processing chips are comprehensively needed with a full steam ahead for the forthcoming V2X system.

\subsection{The Challenging Issues}

\subsubsection{The Path Loss and Channel Model}
\begin{figure}[ht]
\centering
\graphicspath{{figure/}}
\includegraphics[width=3.4in]{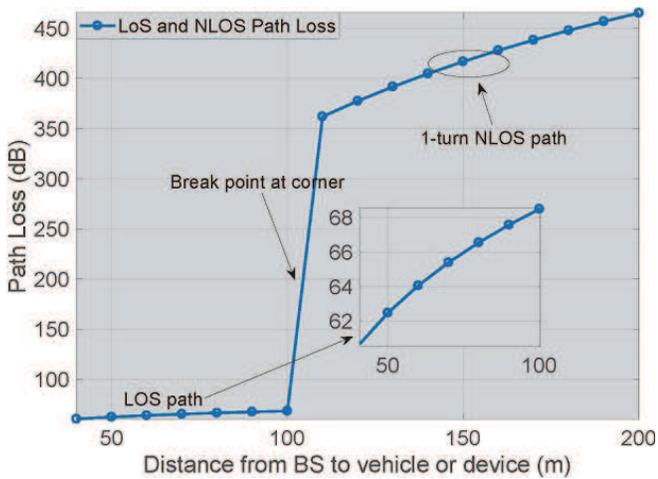}
\caption{The urban path loss simulation according to \cite{Lee15} with BS height 40 m, vehicle and device height 1 m, street block distance 100 m, carrier frequency 2 GHz.}
\label{pl}
\end{figure}

The path loss and channel model of higher carrier frequencies with DSCR and mmWave are vital for cellular coverage and performance estimation, as well as the public and traffic safety. The enriched buildings around, denser BS deployment and even higher carrier frequencies are all challenging issues of the realistic channel and path loss models. In prior studies from UT-Austin, the Manhattan poisson line process was introduced. They assumed the vertical (North-South direction) and horizon (East-West direction) paths growing infinity with $y$ and $x$-axis, based on the street canyons. However, in the non light of sight (NLOS) path, the corner loss was simplified with a constant factor. The ITU-R, recently, published the urban path loss model for the frequency ranges from 300 MHz to 100 GHz. However, it is too complex especially for the mmWave frequency. The light of sight (LOS) and NLOS path losses were comprehensively studied with a remedy study in \cite{Lee15}, the proposed models were verified in Seoul City. We invoke this work here with urban street configuration (street block 100 m, BS height 40 m, vehicle and device height 1 m), the results are given by Fig. \ref{pl}. As shown, NLOS path loss becomes extremely large after 1-turn while crossing the corner. Thus normally, only 1-turn NLOS path can be established in urban street canyon environment. On the other hand, to the mmWave path loss model, series trials were done by the New York University in the New York urban street. Yet up to now, no matured path loss or channel models have been arrived.

\subsubsection{The Low Latency Communication}

In the distributed cache enabled V2X networks, less than 1 ms latency is a vital issue for automatic pilot, especially while encountering some emergency conditions. For instance, giving a vehicle speed 120 Km/h, with LTE latency 30 ms, the moving distance will be around 9 dm. On the contrary, with 5G's 1 ms latency, it will be around 3 cm. Moreover, the low latency is a nature feature for real time in-car entertainment. As known, the ultra dense cellular deployment will result in a limited coverage area, low latency will guarantee the fast response for each request. The handover time across cellular can be reduced as well via low latency. The CoMP joint transmission is calling for low latency processing and precise time synchronization as well.

Albeit 1 ms latency issue has been appealed a lot, in literature, specific technology to achieve the less than 1 ms latency is still less. In our distributed cache enabled V2X networks, although the low latency requirement can be partially realized via D2D communications, and the NOMA\cite{Zhang17} transmission can be invoked for simultaneous transmission of multiple user information, it is still not enough. Generally, to achieve this goal, a dedicated network slicing service should be assigned. The dedicated short package emergency information can be used for fast response to further improve its latency performance. Developing on the chips with real time processing ability is also needed. To sum up, the low latency goal can not be simply achieved, a comprehensive scope on cross-layer design and updates from both hardware and software sides are needed.

\subsubsection{The Ultra Reliable Communication Issue} The ultra reliable communication is another challenging issue. The higher transmission speed will be of less meaning without small error probability. For instance, we prefer 100 MBit/s 99\% reliability rather than 1 GBit/s 50\% reliability while accessing the internet. Ultra reliable communication is critical for vehicle safety and public safety as well. The vehicle cannot be safely braked without it while encountering some emergencies even the processing time is within 1 ms. Additionally, it will cause great damage once the communication link is damaged or hacked, in which case the intelligence control center and automatics pilot cannot successfully manage the traffic.

In literature, the ultra reliable communication is not a new problem. For example, intensive studies have been done in the core network with backup routing and link connections across core routers to ensure the robust connections. For the wireless communication, re-transmission can improve the transmission success rate and the completeness of the received information. However, the massive connected devices bring in new challenging issues for the ultra reliable communication, e.g., hacker attack, resource competing, equipment failure, uncontrollable interference, missing protocol, etc. Additionally, the trade off strategy should be set forth while invoking the re-transmission method with regard to the ultra reliable and low latency requirements. To this end, a lot of work is still needed in future studies.

\section{Conclusion}
The distributed cache enabled V2X networks are introduced in this article. The basic principles, merits, future research trends, and potential challenging issues are discussed. Compared with prior studies, this proposal can offer more intelligent traffic control while connecting all vehicles and devices on the road. The wireless communications are leveraged as well by this proposed system with less engaged transceiver components, better system EE performance and more effective carrier frequency resource usage. More endeavors on the solutions for the even higher transmission rate requirement, flexible network slicing service design and big data/ML/fog computing image recognition and reconstruction methods are needed in future studies. Meanwhile, the accurate urban channel/path loss model, ultra reliable communications and low latency issues are critical factors for its realization.

\begin{bibliographystyle}{IEEEtran}
\begin{bibliography}{IEEEabrv,bibtex}
\end{bibliography}
\end{bibliographystyle}

\end{document}